\documentclass[sigconf,nonacm]{acmart}
\AtBeginDocument{%
  }

\renewcommand\footnotetextcopyrightpermission[1]{}
\setcopyright{none}

\AtBeginDocument{%
  }

\usepackage{subcaption}

\usepackage{xspace}
\makeatletter
\DeclareRobustCommand\onedot{\futurelet\@let@token\@onedot}
\def\@onedot{\ifx\@let@token.\else.\null\fi\xspace}
\def\eg{\emph{e.g}\onedot}

\begin{document}

\title{WISE: A Multimodal Search Engine for Visual Scenes, Audio, Objects, Faces, Speech, and Metadata}

\author{Prasanna Sridhar}
\orcid{0009-0000-5818-1367}
\email{prasanna@robots.ox.ac.uk}
\affiliation{%
  \department{Engineering Science}
  \institution{University of Oxford}
  \city{Oxford}
  \country{UK}
}

\author{Horace Lee}
\orcid{0009-0008-5521-2875}
\email{horacelee@robots.ox.ac.uk}
\affiliation{%
  \department{Engineering Science}
  \institution{University of Oxford}
  \city{Oxford}
  \country{UK}
}

\author{David M. S. Pinto}
\orcid{0000-0003-2710-0186}
\email{pinto@robots.ox.ac.uk}
\affiliation{%
  \department{Engineering Science}
  \institution{University of Oxford}
  \city{Oxford}
  \country{UK}
}

\author{Andrew Zisserman}
\orcid{0000-0002-8945-8573}
\email{az@robots.ox.ac.uk}
\affiliation{%
  \department{Engineering Science}
  \institution{University of Oxford}
  \city{Oxford}
  \country{UK}
}

\author{Abhishek Dutta}
\orcid{0000-0002-5455-3343}
\email{adutta@robots.ox.ac.uk}
\affiliation{%
  \department{Engineering Science}
  \institution{University of Oxford}
  \city{Oxford}
  \country{UK}
}

\begin{abstract}
In this paper, we present WISE, an open-source audiovisual search engine which integrates a range of multimodal retrieval capabilities into a single practical tool, accessible to users without machine learning expertise. WISE supports natural-language and reverse-image queries at both the scene level (\eg empty street) and object level (\eg horse) across images and videos; face-based search for specific individuals; audio retrieval of acoustic events using text (\eg wood creak) or an audio file; search over automatically transcribed speech; and filtering by user-provided metadata. Rich insights can be obtained by combining queries across modalities --- for example, retrieving German trains from a historical archive by applying the object query ``train'' and the metadata query ``Germany'', or searching for a face in a place. By employing vector search techniques, WISE can scale to support efficient retrieval over millions of images or thousands of hours of video. Its modular architecture facilitates the integration of new audio or visual models. WISE can be deployed locally for private or sensitive collections, and has been applied to a number of disparate real-world use cases. Code is available at \underline{\url{https://gitlab.com/vgg/wise/wise}}.
\end{abstract}



\begin{teaserfigure}
  \centering
  \includegraphics[width=0.9\textwidth]{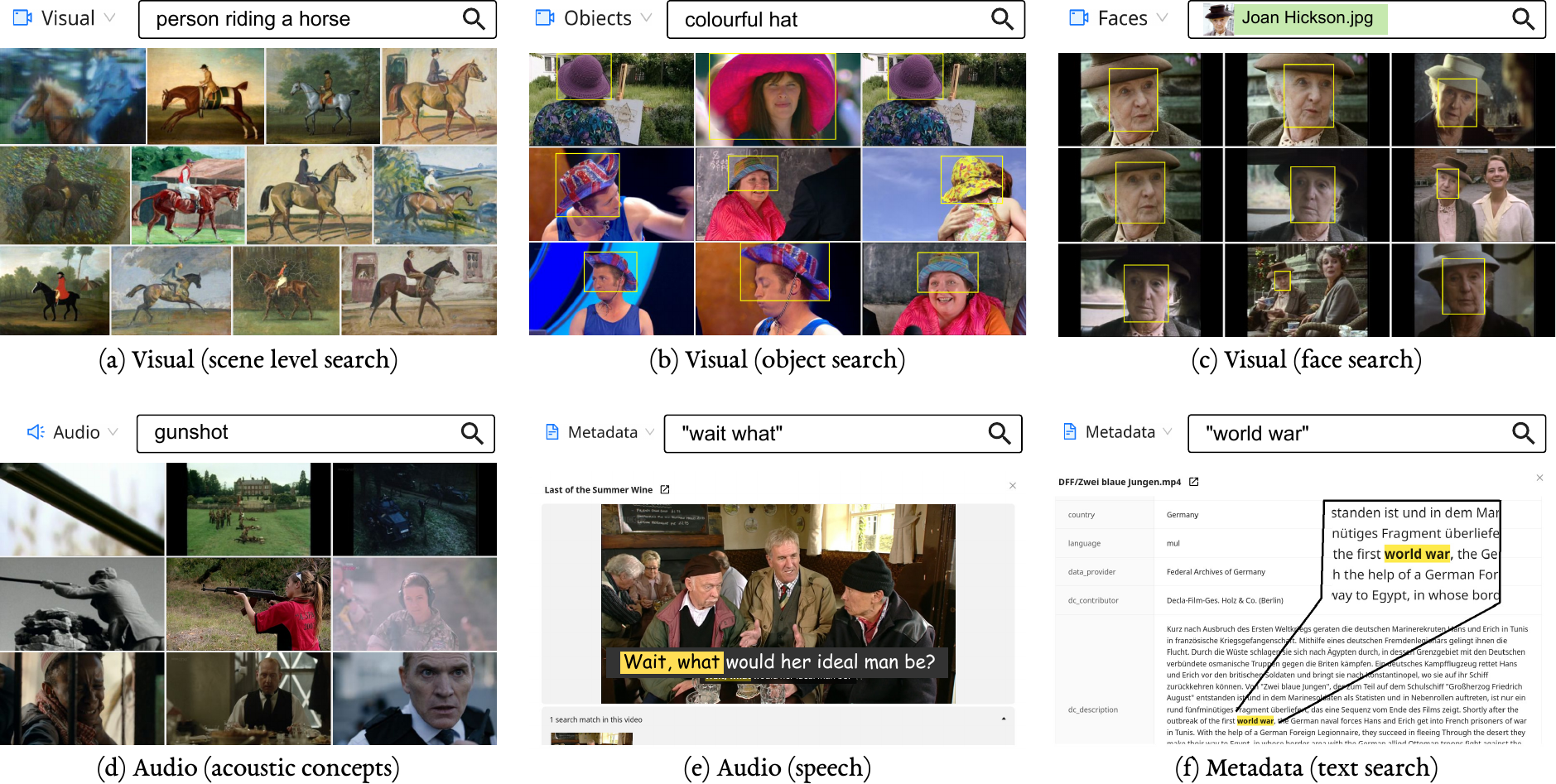}
  \caption{WISE enables search across visual, audio, and metadata streams. Visual search operates at both the scene level (\eg person riding a horse) and the object level (\eg colourful hat) with support for finding specific faces (\eg using a photo of an actress). Audio search retrieves audiovisual segments based solely on acoustic content (\eg gunshot) or speech (\eg ``wait what''). WISE also supports traditional text-based search over media-level metadata.}
  \Description{A teaser image showing search results from various modalities (\eg visual, audio, text) using WISE}
  \label{fig:teaser}
\end{teaserfigure}

\maketitle

\section{Introduction}
Large-scale audiovisual collections are now everywhere, powered by a new wave of devices (\eg mobile phones) and online platforms (\eg social networks) that make it easy to capture, store and share rich multimedia content — images, audio and video - on a massive scale. Although automatically generated metadata (\eg location, date) and manually annotated metadata (\eg title, description, tags) help users explore these collections, they are often limited and cannot fully describe the content. Moreover, collecting high-quality metadata becomes expensive and difficult at scale. In order to be useful, such large collections need search tools that can go beyond metadata and leverage the underlying visual and audio content. While recent advances in multimodal models offer powerful capabilities for content discovery, they remain largely inaccessible to non-technical users due to the expertise required for deployment and integration. This paper introduces WISE (WISE Search Engine), a user-friendly tool which supports searching across visual, audio, and metadata information streams of an audiovisual collection. WISE also offers composite multimodal search capability (\eg metadata $+$ visual or face $+$ visual) as described in Section~\ref{sec:multimodal_search}. The workflow adopted by WISE for processing audiovisual data is defined in Section~\ref{sec:data_processing_and_search} and the software architecture is detailed in Section~\ref{sec:wise_software}. The case studies described in Section~\ref{sec:case_studies} show the impact of the WISE open source software on various research disciplines and industrial sectors. Audiovisual search engines are essential for curating, searching and managing large collections of multimedia content that have become generally available in most research and commercial avenues. The WISE software introduced in this paper has the potential to become a platform for all general purpose audiovisual search requirements.

\section{Multimodal Search Capabilities}

\subsubsection*{Visual Search} The visual stream consists of images and frames extracted from videos. WISE supports visual search at two levels. The first is scene level search, which uses queries that describe the overall composition of a scene. For example, the natural language query ``person riding a horse'' applied to a collection of paintings retrieves images depicting this scene, as shown in \figurename~\ref{fig:teaser}a. An image can also be used as a search query to find visually similar scenes. The second is object-level search, which provides a more fine grained view of the objects within a scene. For example, the query ``colourful hat'' retrieves video segments or images containing hats of various colours and highlights the matched regions with bounding boxes as shown in \figurename~\ref{fig:teaser}b. WISE also supports face search, which can be viewed as a special class of object search, where the results are retrieved by matching the unique identity in facial imagery. \figurename~\ref{fig:teaser}c shows an example using an image of an actor's face as the query.

\subsubsection*{Audio Search} The audio stream in video or audio files can also be searched in WISE using keywords that represent acoustic concepts such as gunshot, laughter, siren, or footsteps. For example, the query ``gunshot'' retrieves video segments or audio files containing the sound of a gunshot, as shown in \figurename~\ref{fig:teaser}d. Although some retrieved segments may also display visual depictions of guns, the results are based solely on the audio content. An audio file can also be used as a search query for audio search. The audio stream frequently includes human speech, and WISE enables search over spoken words using text queries. WISE uses Automatic Speech Recognition (ASR) to transcribe speech into metadata, allowing queries such as ``wait what'' to retrieve relevant video segments where those words are spoken, as illustrated in \figurename~\ref{fig:teaser}e.

\subsubsection*{Metadata Search} Certain metadata (\eg date, location) are typically added to media files automatically while other descriptive metadata (\eg image captions and tags) often need to be added manually by human annotators. WISE supports full text search over this media-level metadata like traditional text-based search tools. For example, the query ``world war'' retrieves all media files whose metadata contains those terms, as shown in \figurename~\ref{fig:teaser}f.

\subsubsection*{Composite Search}
\label{sec:multimodal_search}
\begin{figure}
    \centering
    \includegraphics[width=1\linewidth]{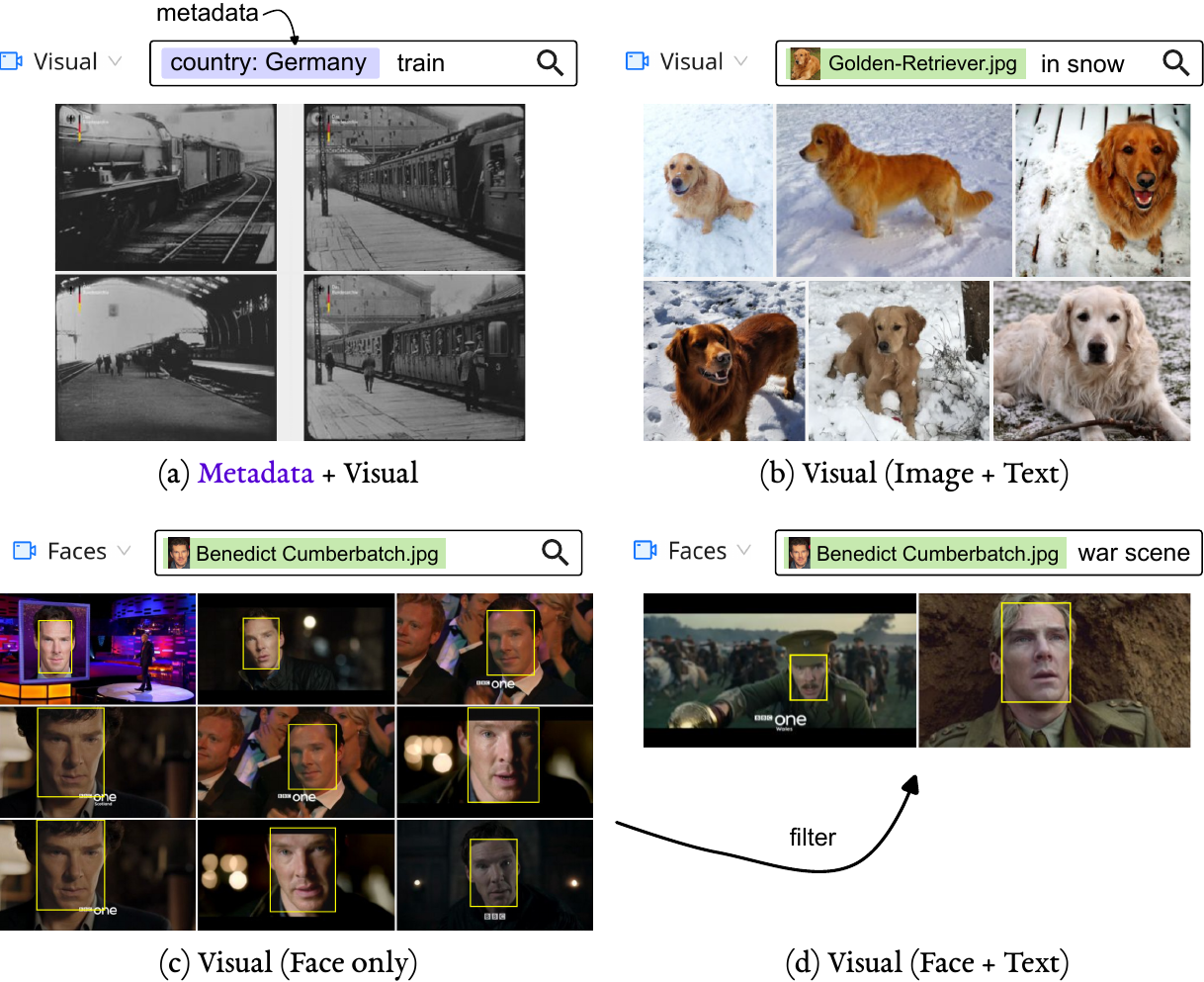}
    \caption{WISE supports multimodal search which allows (a) filtering visual search results using metadata, or (b) performing composed queries by combining an image with a refining text description. Similarly, a compositional query can be applied to face search results (c) to retrieve (d) an actor in a particular type of scene, for example.}
    \Description{A 2x2 grid of images showing result of multi-modal queries in WISE.}
    \label{fig:multimodal-search-collage}
\end{figure}

\begin{figure*}[hbtp]
    \centering
    \includegraphics[width=\linewidth]{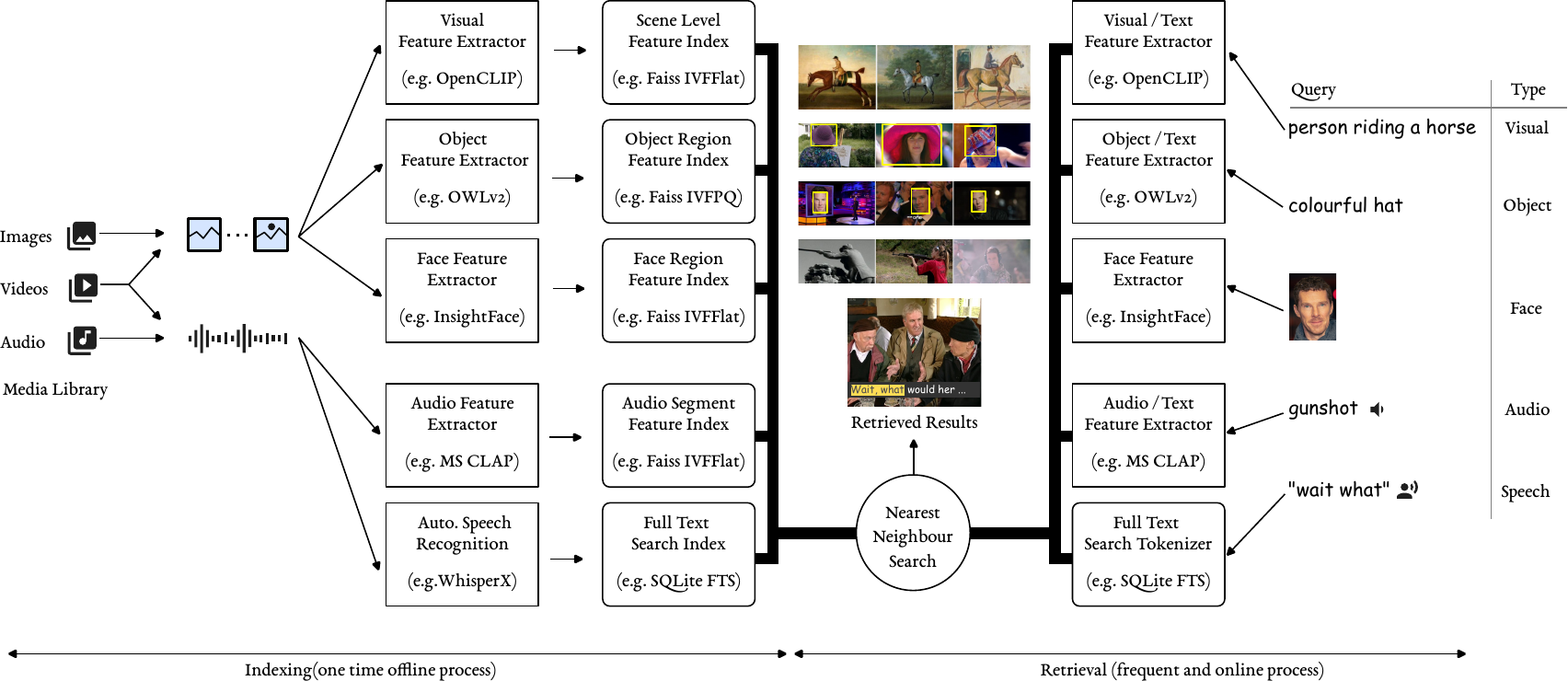}
    \caption{WISE organises the media library into two streams: visual and audio. Scene and region level features (or embeddings) are extracted from the visual stream, which includes images and video frames, while acoustic event and speech features are extracted from the audio stream. All features are saved in a vector store to enable fast nearest neighbour search. Feature extraction and indexing are performed once as an offline process. At query time, audiovisual features are extracted from the search query and matched against the index to retrieve semantically similar content from the media library.}
    \Description{A block diagram showing the data processing workflow implemented by WISE to create a search engine from a set of media files}
    \label{fig:wise-data-processing-block-diagram}
\end{figure*}

WISE supports composite multimodal search by allowing users to combine search queries across modalities. For example, the query in \figurename~\ref{fig:multimodal-search-collage}a applies the metadata filter ``\texttt{country:Germany}'' to a visual search with the natural language query ``train''. This retrieves video segments that contain a train and are tagged with "Germany" in the \texttt{country} metadata field. In many cases, queries are most naturally defined using both an image and a refining text description. This approach, known as composed retrieval~\cite{huynh2025collm}, is illustrated in \figurename~\ref{fig:multimodal-search-collage}b, where an image of a dog combined with the text ``in snow'' retrieves images of dogs in snowy scenes. Face search results, as shown in \figurename~\ref{fig:multimodal-search-collage}c, can also be combined with visual search, enabling queries such as finding specific actors in a particular scene or location, as shown in \figurename~\ref{fig:multimodal-search-collage}d.

\section{Audiovisual Data Processing and Search}
\label{sec:data_processing_and_search}

As shown in \figurename~\ref{fig:wise-data-processing-block-diagram}~(left), WISE aggregates a media library into two streams: visual and audio. The visual stream consists of images and video frames sampled, for example, at $2$ frames per second. The audio stream consists of audio snippets extracted from audio files and audio channel of videos using a window based sampling strategy (\eg 4\,s window with 2\,s overlap). A set of visual feature extractors operate on the visual stream to extract features that can support content-based search at the scene level and region level (\eg object, face). For example, OpenCLIP~\cite{ilharco2025openclip,radford2021learning} features are extracted from each sampled frame to support scene level visual search. Spatial region features are extracted by OWLv2~\cite{minderer2023scaling} and InsightFace~\cite{deng2018arcface} feature extractors to enable visual search based on objects and faces respectively. The scene level features provide a high level overview while the spatial region features offer a more detailed view of the visual stream. Similarly, audio feature extractors operate on the audio stream to extract features that support search for acoustic concepts and speech. For acoustic concepts (\eg gunshot, footsteps, clapping), CLAP~\cite{elizalde2022clap} features are extracted from each window. ASR models (\eg WhisperX~\cite{bain2023whisperx}) are applied to transcribe speech, enabling text-based search of spoken words. 

The extracted features (\eg $768$ dimensional vectors) are stored in a vector search index (\eg Faiss IndexIVFFlat ~\cite{douze2024faiss}) that supports fast approximate nearest neighbour retrieval. For feature extractors that generate a large number of candidate regions (\eg OWLv2), a more aggressive compression scheme (\eg Faiss IndexIVFPQ) can be used to reduce compute and storage costs with a minor impact on retrieval accuracy. The relationship between each feature vector and its associated media content (\eg filename, timestamp, region coordinates) are stored in a SQLite database. Metadata associated with each media file (\eg caption, title) is indexed using traditional full text search (\eg FTS search in SQLite~\cite{sqlite}). Feature extraction and indexing are compute and storage intensive processes, but they only need to be performed once and can be executed offline.

The audiovisual feature extractors used by WISE map text, images, video frames, and audio into a shared vector space. In this space, audiovisual exemplars and their corresponding text descriptions lie close together, while unrelated descriptions are farther apart. These feature extractors are also called vision-language models or audio-language models, and they are trained on a large corpus of audiovisual exemplars and their corresponding text descriptions sourced from the internet. As shown in \figurename~\ref{fig:wise-data-processing-block-diagram} (right), search queries (image or text) are transformed into feature vectors using the same set of feature extractors that were used during indexing. An approximate nearest neighbour algorithm retrieves indexed feature vectors that lie closest to the query vector thereby retrieving the temporal segment or spatial regions in the media library that are most semantically relevant. The retrieval step runs online and completes almost instantly, as it requires only one feature extraction and a fast nearest neighbour search.

\section{Software Architecture}
\label{sec:wise_software}
\begin{figure}[bp]
    \centering
    \includegraphics[width=1\linewidth]{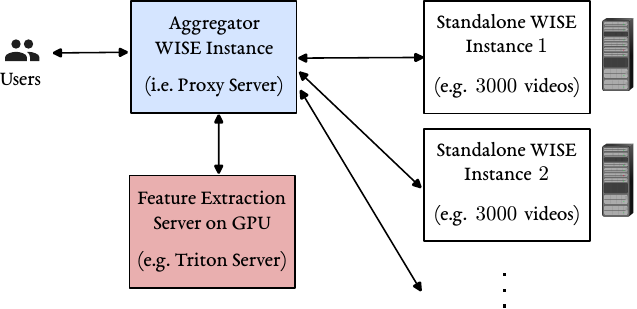}
    \caption{WISE can operate in aggregator mode, where a large audiovisual collection is split up across multiple standalone nodes each covering a different subset. A central instance distributes search queries to all standalone nodes, gathers their results, and presents a unified response to the user.}
    
    \label{fig:wise-aggregator}
    \Description{A block diagram showing the components of the WISE software and how it works in the aggregator mode}
\end{figure}

WISE follows a modular design consisting of five components: Loader, Extractor, Store, Index, and Search. The \textit{Loader} reads media files and extracts audiovisual information in a format (\eg tensor) that enables the \textit{Extractor} to compute feature vectors which are persisted by the \textit{Store}. The \textit{Index} builds vector indexes from stored features to support fast nearest neighbour retrieval. This modular structure provides extensibility, scalability, and high performance.

\subsubsection*{Extensible}
The Loader module can be extended to ingest new media formats (\eg DICOM format from Ultrasound devices). New feature extractors can be integrated by implementing the Extractor's standard interface. For example, replacing the human face feature extractor with ChimpUFE~\cite{iashin2025self} enables the creation of a search engine to study the behaviour and social network of chimpanzees~\cite{bain2021automated}. The Store and Index can also be extended to use alternative storage and indexing backends. WISE's web frontend, built with React, supports straightforward extension of the user interface components.

\subsubsection*{Scalable}
WISE has been deployed on collections as large as 55 million images from Wikimedia Commons~\cite{sridhar2023wise}  and over $6,000$ hours of BBC videos. As shown in \figurename~\ref{fig:wise-aggregator}, WISE can run in \textit{aggregator mode}, distributing queries across multiple machines that each store a subset of the full media collection, and \textit{merging} their results. This architecture allows WISE to scale millions of images and thousands of hours of video.

\subsubsection*{High-performance}
    The Extractor can run on a dedicated GPU machine (see \figurename~\ref{fig:wise-aggregator}), handling batched extraction requests from multiple WISE instances to maximise GPU utilisation. The Loader uses multithreaded pre-fetching to deliver audiovisual data to the Extractor efficiently. A one hour video can be processed in under 10 minutes on a modern computer with a GPU. WISE returns results in under 1 second even for the large datasets aforementioned. Retrieval latency can be reduced further using more efficient indexes such as Product Quantisation with an Inverted File Index~\cite{matsui2018survey}.

\section{Case Studies}
\label{sec:case_studies}

WISE has been adopted in many commercial and academic research projects. Public online demos are available\footnote{\url{https://www.robots.ox.ac.uk/~vgg/software/wise/examples/}} to showcase its deployment in different disciplines. Here are some examples.

\subsubsection*{Journalism}
Open Source Intelligence (OSINT) is a growing trend in journalism which involves developing or investigating a story based on audiovisual content gathered from social media. WISE has been used as part of the reporting and documentary workflow of journalists dealing with a large audiovisual collection, as shown in \figurename~\ref{fig:case-studies-collage}a.

\subsubsection*{Film and Media Research}
\begin{figure}[h]
    \centering
    \includegraphics[width=1\linewidth]{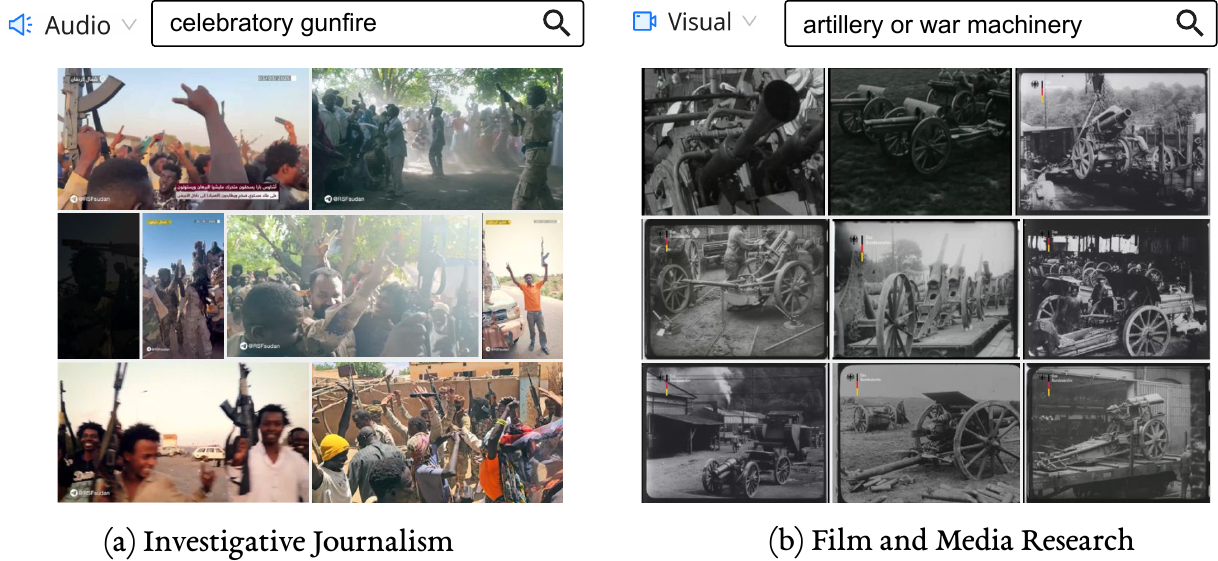}
    \caption{WISE has proved invaluable for journalists investigating stories based on audiovisual content (left) and for film and media scholars studying historical archives (right).}
    \label{fig:case-studies-collage}
    \Description{Two screenshots from the WISE software highlighting the investigative journalism, film and media research use cases. The screenshot for the investigative journalism use case shows the query "celebratory gunfire" against the audio stream and the search results of people waving guns in the air. The screenshot for the film and media research shows the query "artillery or war machinery" and the corresponding results from the historical archives depicting the war equipment}
\end{figure}

WISE supports film and media scholars in curating and exploring historically significant audiovisual collections. An \underline{\href{https://meru.robots.ox.ac.uk/cinephile/}{online demo}} developed for the \underline{\href{https://hermes-hub.de/forschen/datachallenges/challenges/challenge-2025.html}{Cinephile Challenge 2025}} demonstrates its use on material from the Dutch and German national archives, as shown in \figurename~\ref{fig:case-studies-collage}b.

\subsubsection*{Other Use Cases}
Wildlife conservation organisations are using WISE to explore their video archives to identify relevant clips for environmental storytelling. Commercial archives employ WISE to improve retrieval and licensing of historically significant media.

\section{Related Work}
WISE is based on vision-language (\eg \cite{ilharco2025openclip,radford2021learning}) and audio-language (\eg \cite{elizalde2022clap}) models that embed different modalities (\eg visual, audio, text) in a shared embedding space, as described in Section~\ref{sec:data_processing_and_search}. Some commercial entities (\eg OpenAI, Cloud Providers) offer embedding model API services and leave the building of the search functionality and interfaces up to the user. Software-as-a-service multimedia search tools, \eg Google Photos, TwelveLabs, and muse.ai, are more accessible to non-technical users, but often require a paid plan (when usage exceeds free tier limits) and rely on users to upload their data online. Other open source tools such as Exquisitor~\cite{sharma2025exquisitor} and CineSearcher~\cite{kreten2025cinesearcher} also leverage vision-language models to enable natural language search over visual content, while delivering novel search functionalities like conversational search, interactive exploration and an iterative workflow for annotating multimedia content. In comparison, WISE offers search over additional specific visual content (faces, objects), audio and ASR; it enables search queries that combine modalities (\eg a query composed of text and an image); and has a more flexible architecture allowing feature extractors to be easily changed as new and better models become available.

\section{Discussion and Conclusion}
WISE leverages recent advances in vision-language~\cite{ilharco2025openclip} and audio-language~\cite{elizalde2022clap} models to enable exploration of audiovisual collections using text and image based queries. By allowing users to combine and filter search results from multiple modalities, our software has facilitated the use of audiovisual search in domains such as film and media research, and journalism which have so far relied mostly on manual curation and search. With a design that enables usage on a local machine, WISE allows users working with sensitive or proprietary collections to avoid the need to upload content to third party providers on the internet. The system can be extended to incorporate newly developed audiovisual models and search indexes. WISE is available open-source under the Apache License 2.0 and we hope that this work can serve as a practical tool across diverse audiovisual collections.

\begin{acks}
  This work was funded by the
  \grantsponsor{T028572/1}{EPSRC}{https://doi.org/10.13039/501100000266}
  \grantnum{T028572/1}{Programme Grant VisualAI EP/T028572/1}. We are grateful to Dr. Ashish Thandavan for supporting the compute infrastructure requirements of this project, and the reviewers for their comments.
\end{acks}

\bibliographystyle{ACM-Reference-Format}
\balance
\bibliography{ref}
\end{document}